\documentclass[prl,showpacs,twocolumn,amsmath,amssymb]{revtex4}

\usepackage{graphicx}
\usepackage{dcolumn}
\usepackage{bm}

\begin{document}

\title{Anisotropy of the upper critical field in superconductors
with anisotropic gaps. Anisotropy parameters of MgB$_2$}

\author{P. Miranovi\' c}   
\author{K. Machida}
\affiliation{ Department of Physics, Okayama University,
700-8530  Okayama, Japan}
\author{ V. G. Kogan }
\affiliation{ Ames Laboratory DOE and Physics Department ISU,
Ames, IA 50011, USA.}


\begin{abstract}
The upper critical field $H_{c2}$ is evaluated for weakly-coupled
two-band anisotropic  superconductors. By modeling the actual bands and
the gap distribution of MgB$_2$ by two Fermi surface spheroids with
average parameters of the real material, we show that
$H_{c2,ab}/H_{c2,c}$ increases with decreasing temperature
in agreement with available data.
\end{abstract}

\maketitle

The anisotropic Ginzburg-Landau (GL) equations,  derived for clean
superconductors with   arbitrary gap anisotropy by
  Gor'kov and  Melik-Barkhudarov \cite{Gorkov}, led to a common
practice of characterizing materials by a single
anisotropy parameter defined as $\xi_a/\xi_c\equiv
\lambda_c/\lambda_a$ ($\xi$ is the coherence length, $\lambda$ is
the penetration depth, and $a,c$ are principal directions of a
uniaxial crystal of the interest here). Formally, this came out
because in the GL domain, the same ``mass  tensor"   determines
the anisotropy  of both $\xi$ (of the upper critical fields
$H_{c2}$) and of $\lambda$.

At arbitrary temperatures, however, the ratios of $H_{c2}$'s and of
$\lambda$'s are not necessarily the same.
We demonstrate below that in materials with
anisotropic Fermi surfaces and anisotropic gaps, not only
$H_{c2,a}/H_{c2,c} $ may strongly depend on
$T$, but this ratio might differ considerably from
$\lambda_c/\lambda_a$ at low $T$'s. Our arguments are based on the
weak-coupling model of superconductivity for simple
Fermi surfaces and gap anisotropies;  as such they are at best
qualitative. Still,  being applied to MgB$_2$, they provide
satisfactory description of existing data for the $H_{c2}$
anisotropy.

  There are two different approaches in literature in describing
macroscopics of the ``two-gap" superconductivity of MgB$_2$. One of
these   treats the anisotropy of interaction by introducing  a
coupling   matrix $\rho_{ij}$ for intra- and interband pair transfer.
Various relations between the matrix elements yield variety of
macroscopic consequences. One of the realizations of the model
considers two order parameters with two  distinct phases which may
lead to various static \cite{golubov,Ask} and dynamic \cite{GV}
effects. Certain relations between   elements
$\rho_{ij}$ provide  the experimentally observed gaps and their ratio.
The approach adopted in this
paper  considers the gap on two Fermi sheets just as a particular case
of the gap anisotropy, the gap ratio being the experimental input
parameter. Within this scheme, there is only one complex order
parameter $\Psi$, a single critical temperature $T_c$ is built in, and
the number of input parameters needed for calculations is small.   The
resulting $H_{c2}$'s of these two approaches are similar 
\cite{MMKunpublished}. Our choice   is dictated  by theoretical 
simplicity, rather than by
experimental necessity.

Below,  the general scheme of calculating $H_{c2}(T)$ based on the
Eilenberger method \cite{E} is  outlined for both the gap  and the
Fermi surface being anisotropic \cite{RS}. Then, we   estimate
   the $H_{c2}$ anisotropy of MgB$_2$ by modeling the four-sheet
Fermi surface as calculated in Refs. \onlinecite{Bel,Mazin,Choi} with
the gap distribution having two sharp maxima, by two distinct Fermi
sheets $F_{1,2}$ with  gaps $\Delta_{1,2}$, each being constant within
its sheet. Since the actual band structure enters  macroscopic
superconducting parameters via Fermi-surface averages, we further
model the sheets $F_{1,2}$ by two spheroids with average Fermi
velocities close to the band-structure generated values. As a
result we obtain qualitatively (and - given the spread of existing
data - quantitatively) correct behaviors of
$H_{c2}(T)$ for both principal directions. \\

A clean weakly coupled superconductor at $H_{c2}(T)$ is described within
the quasiclassical Eilenberger scheme by the anomalous Green's function
$f(\omega,{\bm v},{\bm r})$, which satisfies an equation:
  \begin{equation}
  (2\omega+\bm v \bm\Pi )f  =
2\Delta /\hbar\,. \label{eil1}
\end{equation}
Here, $\bm v$ is Fermi velocity; $\omega=\pi T(2n+1)/\hbar$ are the
Matsubara frequencies;
$\bm \Pi=\bm\nabla+(2\pi i/\phi_0)\bm A$ with the vector
potential $\bm A$  and the flux quantum $\phi_0$.
  The Eilenberger function $g$ describing   normal excitations, is
unity at $H_{c2}(T)$. Equation (\ref{eil1}) holds for any Fermi surface
and gap anisotropies.

  The gap function $\Delta$ satisfies the self-consistency equation:
\begin{equation}
\Delta({\bf r},{\bm v})=2\pi TN(0) \sum_{\omega >0}^{\omega_D} \left<
V({\bm v},{\bm v}^{\prime\,}) f(\omega,{\bm v}^{\prime\,},{\bf r})
\right>_{{\bm v}^{\prime\,}}\,.   \label{eil4}
\end{equation}
Here, $\left<...\right>$ denotes the average over the Fermi surface.
The pair scattering potential $V$ is assumed factorizable,
$ V({\bm v},{\bm v}^{\prime\,})=V_0 \,\Omega({\bm v})\,\Omega({\bm 
v}^{\prime\,})$, with the function $\Omega({\bm v})$   normalized so
that
\begin{equation}
  \left< \Omega^2 \right>=1\,.
\label{norm}
\end{equation}
Further, we look  for   $\Delta (
{\bf r},T;{\bm v})=\Psi ({\bf r},T)\, \Omega({\bm v})$ \cite{Pokr},
the form implying a one-component order parameter. Then,  after
excluding
$V_0$ with the help of the  BCS formula for
$T_c$, one arrives to
   Eq. (\ref{eil4}) of the form:
  \begin{equation}
\frac{\Psi }{2\pi T} \ln \frac{T_{c}}{T}= \sum_{\omega>0}^{\infty}
\Big(\frac{\Psi}{\hbar\omega}- \left <  \Omega \, f  \right > \Big)\,
\label{gap1}
\end{equation}
(for more details and references see, e.g., Ref.
\onlinecite{RapCom}).

The linear Eq. (\ref{eil1}) is inverted:
\begin{equation}
f =\frac{2}{\hbar}\left(2\omega + \bm v \bm\Pi\right)^{-1}\Delta =
\frac{2\Omega}{\hbar}\int\limits_0^\infty d\rho \,
e^{-\rho (2\omega + \bm v \bm\Pi )}\Psi\,.
\end{equation}

In the GL domain, the gradients $\Pi\sim \xi^{-1}\to 0$, and one can
keep in the expansion of  $\exp(-\rho{\bm v}  {\bm
\Pi})$ only the terms up to the second order. Then, Eq. (\ref{gap1})
yields:
\begin{equation}
  -\Psi \delta t=  \frac{7\zeta (3)\hbar^2}{16\pi^2T_c^2}\,
   \left<\Omega^2 ({\bm v}  {\bm
\Pi})^2\Psi\right> \,,  \label{GL'}
\end{equation}
where $\delta t=1-T/T_c$. This is  the anisotropic  linearized GL
equation of Ref. \onlinecite{Gorkov}:
\begin{equation}
- \xi^2_{ik}  \Pi_i\Pi_k \Psi =\Psi\,,\quad  \xi^2_{ik}  = 
\frac{7\zeta (3)\hbar^2}{16\pi^2T_c^2 \delta t}\,
   \left<\Omega^2   v_iv_k \right> \,.  \label{GL}
\end{equation}
  The anisotropy parameter in this domain follows:
\begin{equation}
\gamma_H^2(T_c) = \frac{H_{c2,a}^2}{H_{c2,c}^2} =
\frac{\xi_{aa}^2}{\xi_{cc}^2}=
\frac{\langle \Omega^2 v_a^2\rangle }{\left< \Omega^2 v_c^2\right>}\,.
\label{anis_clean}
\end{equation}
It should be stressed that at $T_c$, the anisotropy of the London
penetration depth $\gamma_{\lambda}(T_c) =\lambda_{c}/\lambda_a$ is
determined by the same tensor $\left<\Omega^2   v_iv_k \right>$,
i.e.,  $\gamma_{\lambda}(T_c) =\gamma_H(T_c)$, see Refs.
\onlinecite{Gorkov} or \onlinecite{RapCom}.  \\

{\it The formal method.}
To treat arbitrary $T$'s, we follow Helfand-Werthamer's routine
\cite{HW} (for Eilenberger based procedure see, e.g., Refs.
\onlinecite{RS,K88}): introduce
$v^{\pm}=v_x\pm iv_y$ and
$\Pi^{\pm}=\Pi_x\pm i\Pi_y$ for the field along $z$,  write
$\bm v \bm\Pi= \left( v^{ +}\Pi^-+v^{ -}\Pi^+\right)/2$, and use
known properties of exponential operators to evaluate $\exp(-\rho \bm
v \bm\Pi)$.  There are a few ways to proceed with actual calculation.
E.g., by  writing $\omega^{-1}=2\int_0^{\infty}d\rho \,e^{-2\rho
\omega}$, one can sum up over $\omega$ in Eq. (\ref{gap1}) and
evaluate numerically the remaining integrals over $\rho$. We found it
a faster procedure to follow Rieck and Scharnberg \cite{RS}: first to
express integrals over $\rho$ in terms of Repeated Integrals of the
Error Function ${\bf i}^n {\rm erfc}(x)$ \cite{Abr}, and then
numerically evaluate convergent sums over $\omega$.  We then obtain:
\begin{eqnarray}
&&f =\frac{\Omega}{\hbar
}\sqrt{\frac{2\phi_0}{H}}\sum\limits_{n,m=0}^\infty
I_{nm}\,(\hat{a}^+)^m(\hat{a}^{-})^n\Psi \,,\label{Inm} \\
&&I_{nm}=
\frac{(n+m)!}{n!m!} \left(i\sqrt{2} \right)^{n+m}
\frac{  (v^-)^{m-n}}{v_{\perp}^{m-n+1}}\,
  {\bf i}^{n+m}\mathrm{erfc}\left(x
\right) e^{x^2},\nonumber
  \end{eqnarray}
where $\hat{a}^{\pm}=i\sqrt{\phi_0/4\pi H  }\,\Pi^{\pm}$,
$x=(\omega/ v_{\perp})\sqrt{2\phi_0/\pi H}$ and
$v_{\perp}^2=v_x^2+v_y^2$.

To satisfy the self-consistency equation we expand $\Psi$ in a complete
set of functions $\Psi_j$ constructed by Eilenberger to
represent various vortex lattice solutions:
\begin{equation}
\Psi=\sum_{j=0}^{\infty}C_j\Psi_j \,. \label{expansion}
\end{equation}
In fact, we do not need the explicit  form of $\Psi_j$ (they can
be found in Ref.  \onlinecite{Eil}); suffices it to mention that
$\Psi_0$ has the known Abrikosov form, whereas the rest are found
by applying $\hat{a}^+$ to $\Psi_0$:
$\hat{a}^+\Psi_n=\sqrt{n+1}\Psi_{n+1}$ and
$\hat{a}^-\Psi_n=\sqrt{n }\Psi_{n-1}$.
After some algebra, we arrive at the following equation:
\begin{eqnarray}
&
\displaystyle
\sum\limits_{\omega>0}\sum\limits_{j,m}^{\infty}\sum\limits_{n=0}^j
\left<\Omega^2I_{nm}\right>
\dfrac{\sqrt{j!(j-n+m)!}}{(j-n)!}\;C_{j}\Psi_{j-n+m}=&\nonumber\\
&\displaystyle
\frac{\hbar \sqrt{H}}{2\pi T\sqrt{2\Phi_0}}\left(
\ln\frac{T}{T_c}+2\pi T\sum\limits_{\omega>0}\frac{1}{\hbar\omega}
\right)\sum\limits_{j=0}^{\infty} C_{j}\Psi_j\,.&
\label{hc2}
\end{eqnarray}

As argued in Ref. \onlinecite{RS}, the integers $j$ and $n-m$ should be
taken even. When  the coefficients in front of each $\Psi_k$ are
set  zero, the homogeneous  system of linear equations for
$C_k$ is obtained. The determinant of this system has to be zero,
which gives an equation for $H$; the highest root is the upper critical
field.\\

{\it MgB$_2$.} The band structure calculations show that   the
Fermi surface of this material consists of sheets coming from two
$\pi$-bands  and   two quasi-two-dimensional $\sigma$-bands
\cite{Bel,Mazin,Choi}. It has been shown by solving the Eliashberg
equations \cite{Choi,Mazin} that the gap on the
four Fermi surface sheets   has two sharp maxima:
$\Delta_1\approx 1.7\,$meV at the two   $\pi$-bands  and
$\Delta_2\approx 7\,$meV at the two $\sigma$-bands. Within each of
these groups, the spread of the gap values is small, and the gaps can be
considered as constants, the ratio of which is nearly $T$ independent.
Experimental estimates of the mean-free path $\ell$ show that in most
of available samples $\ell\gg\xi(0)$, i.e., the material can be
considered as a clean  superconductor.

The macroscopic anisotropies such as those of $H_{c2}$, of the
penetration depth $\lambda$, of the magnetization, etc., are determined
within a weak-coupling theory by the Fermi surface averages [see, e.g.,
Eqs. (\ref{GL})  and (\ref{anis_clean})]. Therefore, for qualitative
study of these anisotropies, fine details of the Fermi surfaces are
unlikely to be relevant.   We then {\it model} the Fermi surface
of this material as two separate sheets $F_1$ and $F_2$ with average
characteristics taken from actual band structure results.  Thus, we
   consider a model material with the gap anisotropy given by
\begin{equation}
\Omega ({\bm v})= \Omega_{1,2}\,,\quad {\bm v}\in   F_{1,2} \,.
  \label{e50}
\end{equation}
Denoting the densities of states on the two parts as $N_{1,2}$,  we
obtain for the general averaging:
\begin{equation}
\left< X \right> =  \frac{N_1\left<X_1\right>+
N_2\left<X_2\right>}{N(0)} =
\nu_1\left<X_1\right>+ \nu_2\left<X_2\right>\,,\label{norm2}
\end{equation}
where $N(0)$ is the total density of states and we introduced
normalized densities of state $\nu_{1,2} $ for brevity.  We have then
instead of Eq. (\ref{norm}):
\begin{equation}
\Omega_1^2 \nu_1+\Omega_2^2\nu_2=1\,,\quad \nu_1+\nu_2=1\,.\label{norm1}
\end{equation}
   We also assume that the two parts of the Fermi surface have the
symmetries of the total, e.g., $\left< {\bm v}\right>_{1}=0$ where the
average is performed only over the first Fermi sheet.

According to Refs. \onlinecite{Bel,Choi}, the relative densities of
states
$\nu_1$ and $\nu_2$ of our model are $\approx\,$0.56 and 0.44.
  The ratio $\Delta_2/\Delta_1=\Omega_2/\Omega_1\approx
4$, if one takes the averages of $6.8\,$ and $1.7\,$meV
for the two groups of distributed gaps as calculated in Ref.
\onlinecite{Choi}. Then, the normalization (\ref{norm1})  yields
$\Omega_1=0.36$  and  $\Omega_2=1.45$. One should have in mind that
the data on the gaps ratio vary from less than 3 (see, e.g.,
\cite{Bouquet} and references therein) to more then 4 \cite{Badr}.

In the following we  use   averages over separate Fermi sheets
calculated in Ref. \onlinecite{Bel}: $\left< v_a^2\right>_1=33.2$,
$\left< v_c^2\right>_1=42.2$,  $\left< v_a^2\right>_2=23$,
  and $\left< v_c^2\right>_2=0.5\times 10^{14}\,$cm$^2$/s$^2$. It is
worth noting that  the ratio
$\left< v_a^2\right>/\left< v_c^2\right>$ is   0.79 for $F_1$, and
   46 for $F_2$, whereas for  the whole Fermi surface it is 1.2\,.

We further model the pieces $F_{1,2}$ by spheroids, in other words, we
consider quadratic energy spectra $E_\mu(\bm
k)=(k_x^2+k_y^2)/(2m_{a,\mu })+ k_z^2/(2m_{c,\mu })$,
$\mu=1,2$. Within this simple model, evaluation of averages is tedious
yet straightforward;  to get the averages given above we need
$m_{a,1}/m_{c,1}=1.3$ and $m_{a,2}/m_{c,2}=0.029$.

\begin{figure}[t]
\includegraphics[angle=0,scale=0.4]{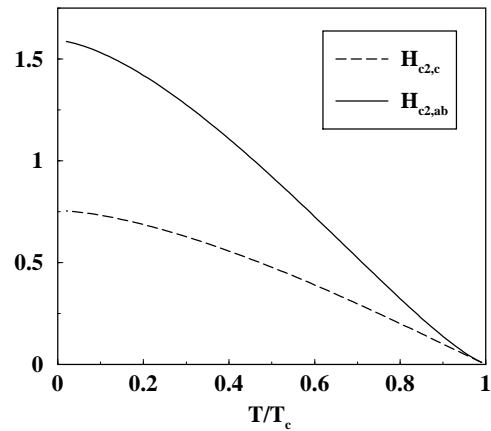}
\caption{Temperature dependence of the upper critical field for
two principal directions. Note: each field  is normalized on its own
slope at $T_c$:  $T_c(\mathrm{d}H_{c2}/\mathrm{d}T)_{T_c}$.}
\label{fig1}
\end{figure}

\begin{figure}[t]
\includegraphics[angle=0,scale=0.4]{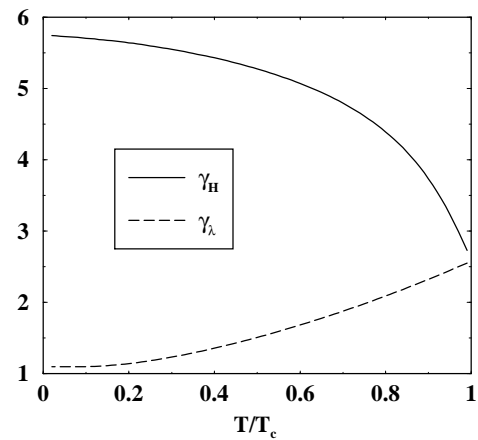}
\caption{
Anisotropy ratio $\gamma_H=H_{c2,ab}/H_{c2,c}$ versus
$ T/T_c$ calculated with parameters for MgB$_2$ given in the text.
Dashed line is $\gamma_\lambda(T)=\lambda_c/\lambda_{ab}$ calculated
as in Ref. \onlinecite{RapCom}.
}
\label{fig2}
\end{figure}

The expansion (\ref{expansion}) converges fast at high $T$'s, but we
need terms  up to $\Psi_{20}$ to stabilize the low temperature values
of $H_{c2,a}$. The temperature dependence of $H_{c2}$  for two
principal directions so calculated is shown in Fig. \ref{fig1}. The
curve   $H_{c2,c}(T)$ is similar to that of
Helfand-Werthamer \cite{HW}, the result confirmed by the data of Ref.
\onlinecite{sologubenko}. On the other hand,
$H_{c2,a} $ has  a slight positive curvature at high temperatures.
As a result, the ratio $H_{c2,a} /H_{c2,c}$ is temperature dependent,
as shown by the upper curve of Fig. \ref{fig2}.

It is of interest to note that within a few \%, solutions $\Psi$ of
the self-consistency  Eq. (\ref{hc2}) can be
approximated by  solutions of
$-\xi_{ik}^2\Pi_i\Pi_k\Psi =\Psi$, with properly chosen parameter
$\gamma_H(T)=\xi_a/\xi_c$.  The reasons for this will be discussed
elsewhere \cite{MMKunpublished}. Physically, this means that the
angular dependence  of $H_{c2}$ at any $T$ should be close to:
\begin{equation}
H_{c2}(\theta,T)=\frac{H_{c2,ab}}{\sqrt{\gamma^2_H(T)\cos^2\theta
+\sin^2\theta}}\,,
\end{equation}
  where $\theta$ is the angle between the applied field and the $c$
axis. This angular dependence has, in fact, been recently reported
\cite{Gren}.

The drop of the  $H_{c2}$ anisotropy with increasing $T$ has been
recorded by Angst {\it et al.} in measurements on single crystals of
MgB$_2$ \cite{Angst}.  Bud'ko and Canfield used a robust method of
extracting the  anisotropy of $H_{c2}$ \cite{BKC} in the whole
temperature range using the $T$ dependence of the magnetization of
random powders \cite{BC}. Recent specific heat measurements of Lyard
{\it et al.} \cite{Gren} and magnetization data of Welp {\it et al.}
\cite{Welp} on single crystals produced similar results.  All these
data show qualitatively similar behavior to that of the upper curve
of Fig. \ref{fig2}.

Physically, the large anisotropy of $H_{c2}$ at low temperatures
($\approx 6$ in our calculation) is related to the large gap value at
the Fermi sheet which is nearly two-dimensional. With increasing $T$,
the thermal mixing with the small-gap states on the three-dimensional
Fermi sheet suppresses the anisotropy down to 2.6 at $T_c$.

It is worth to compare here the  $H_{c2}$ anisotropy with the
anisotropy of the penetration depth
$\gamma_{\lambda}(T)=\lambda_c/\lambda_a$ calculated within the same
model of MgB$_2$ \cite{RapCom}. The result is shown by the dashed
line in Fig. 2. At $T_c$, the $H_{c2}$ anisotropy $\gamma_H$ coincides
with $\gamma_{\lambda}$, as they should because the GL theory contains
only one ``mass tensor" which determines anisotropies of both
$\xi^{-2}$ and $\lambda^2$. The expression (\ref{anis_clean}) for the
anisotropy ratio clearly amplifies the contribution of Fermi surface
parts with the large gap. However, in a clean material at $T=0$, the
superfluid of Cooper pairs is Galilean invariant; this implies that
all charged particles participate in the superflow, their energy
spectrum notwithstanding. This is reflected in the result for the
$\lambda$-anisotropy at $T=0$,
\begin{equation}
\gamma^2_{\lambda}(0)=\frac{\left<v_a^2\right>}{\left<v_c^2\right>}\,,
\label{ratio}
\end{equation}
neither  the gap nor its anisotropy  enter this expression. As
mentioned, for MgB$_2$, the  ratio (\ref{ratio}) is 1.2, i.e.,
$\gamma_{\lambda}(0)=1.1$ in a striking difference with the
large zero-$T$ anisotropy of $H_{c2}$.

Thus, the question ``what is the anisotropy parameter of MgB$_2$?" does
not have a unique answer. To pose the question properly one should
specify the quantity of interest. If this is $H_{c2}$, the answer is
given by the upper curve of Fig. 2 for a clean material; if this is
$\lambda$, see the dashed line. If this is the magnetization in
intermediate fields, $M\propto (\phi_0/\lambda^2)\ln (H_{c2}/H)$, the
main contribution to anisotropy comes from $\lambda$, however, the
$H_{c2}$ anisotropy contributes as well (being smoothed by the
logarithm). The last situation should be taken into account while
extracting anisotropy from the  torque data in tilted fields
\cite{torque1,torque2}, the point stressed by  Angst \cite{private}.

  Given the simplifying assumptions about the Fermi surfaces we have
made and our  results for the $H_{c2}$ anisotropy which reproduce
qualitatively well the measured anisotropy, we believe that the
anisotropic properties of $H_{c2}$ and $\lambda$ we have described,
are  generic for materials with anisotropic gaps and Fermi surfaces.

\end{document}